\documentclass{PoS}

\title{From many body wee parton dynamics to perfect fluid: a standard model for heavy ion collisions}

\ShortTitle{Standard model of heavy ion collisions}

\author{\speaker{Raju Venugopalan}\thanks{This work was supported by the US Department of Energy under DOE Contract No.DE-AC02-98CH10886. I would like to thank Roy Lacey for a careful reading of the manuscript.}\\
        Physics Department, Bldg. 510A, Brookhaven National Laboratory, Upton, NY 11973, USA\\
        E-mail: \email{raju@bnl.gov}}

\abstract{We discuss a standard model of heavy ion collisions that has emerged both from the 
experimental results of the RHIC program and associated theoretical developments. We comment briefly on the impact of early results of the LHC program on this picture. We consider how this standard model of heavy ion collisions could be solidified or falsified in future experiments at RHIC, the LHC and a future Electron-Ion Collider.}

\FullConference{35th International Conference of High Energy Physics - ICHEP2010,\\
		July 22-28, 2010\\
		Paris France}

\begin{document}

\section{Introduction}
Quantum Chromodynamics (QCD) is the ``nearly perfect" theory of the strong interactions~\cite{Wilczek}. The QCD Lagrangean, expressed in terms of fundamental quark and gluon fields, is rich in symmetries and is parameter free if quarks are taken to be massless. Phenomena in the world where the strong interaction holds sway arise as a consequence of the dynamics of the quark and gluon fields. In particular, the QCD vacuum plays an important role in the structure of strongly interacting matter. A subtle example of the role of the vacuum is chiral symmetry breaking, whereby the masses of light quarks are leveraged via the Nambu-Goldstone mechanism into the much heavier masses of the spectrum of observed hadrons. 

Because QCD is a theory where most phenomena are "emergent" phenomena, it is especially interesting to ask what happens to QCD matter when squeezed to ultra-high densities or heated to ultra-high temperatures. 
The role of the vacuum is greatly modified and strongly interacting matter may behave very differently under extreme conditions relative to those found at low temperatures and densities. 

The structure of QCD has immediate consequences for extremely hot and dense matter which resolve puzzles about the strong interactions present since its early days such as, for example, ``limiting temperatures". Because of asymptotic freedom, QCD matter asymptotically must be a deconfined system of weakly interacting quarks and gluons~\cite{Cabibo-Parisi}. Likewise, for matter at extremely large baryon densities, the close packing of hadrons gives way to quark matter~\cite{Collins-Perry}. Because screening of quarks and gluons plays a significant role in the dynamics of QCD matter at high temperatures and densities, such a deconfined state is called a Quark-Gluon Plasma~\cite{Shuryak}. Our universe was a QGP  $\sim 10^{-5}$ seconds after the Big Bang. 

Another key feature of QCD is "infrared slavery" at large separations (or low temperatures and densities)--a phenomenon closely related to confinement. As a consequence, the QGP undergoes a transition  from deconfined quark-gluon matter to confined hadronic matter. Similarly, chiral symmetry which is restored at high temperatures is broken at lower temperatures. It is therefore clear that the theory must contain 
a rich phase diagram for variations in temperature and baryon chemical potential. Significant theoretical developments in lattice gauge theory and an explosion of computing power have made it possible to explore aspects of the phase diagram in great detail. We refer the reader to Owe Philipsen's talk at this conference for details on the latest developments~\cite{Philipsen}.

Our focus here is on heavy ion collisions at ultrarelativistic energies where the aim is to produce deconfined QCD matter in the laboratory--this matter is the hottest and densest matter produced on earth and lasts for $\sim 3\cdot 10^{-23}$ seconds. We will argue here that a  wide range of results from the RHIC experiments coupled with theoretical developments have given rise to a picture of heavy ion collisions on which there is a sufficiently broad consensus for us to describe it as a "standard model" of heavy ion collisions. Formulating a standard model is useful and perhaps essential for an intellectually coherent narrative of the science, to expose clearly possible weak chinks in such  armor and to bring clarity to future paths along which the field could develop.

\section{A standard model of heavy ion collisions}

Exploring the QCD phase diagram at high temperatures and densities in heavy ion collisions is challenging because the collisions are intensely dynamical processes where contributions to any measured final state can arise from different regions in the space-time evolution of the collision. One therefore needs to understand the entire chronology of the collision starting from the properties of the high energy nuclear wavefunctions. While historically the focus of heavy ion collisions was primarily on understanding the thermal properties of deconfined QCD matter, there has been a gradual realization that other aspects of the collision open additional interesting windows into the many-body dynamics of QCD in collision. 

\begin{figure}[h]
\centering
\includegraphics[width=125mm]{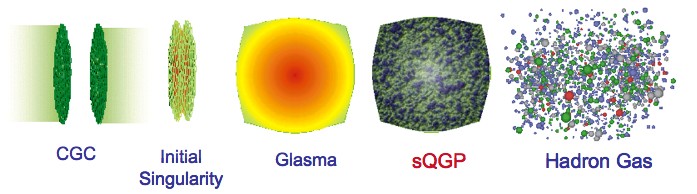}
\caption{A standard model of the space-time evolution of heavy ion collisions.} 
\label{fig:bass}
\end{figure}A standard heavy ion model consists of the chronology shown in fig.~\ref{fig:bass}. 
Before the collision, the nuclear wavefunctions are described by correlated multi-parton states called Color Glass Condensates (CGC). The CGCs shatter in the collision producing deconfined non-equilibrium QCD matter called the Glasma. This Glasma likely thermalizes to form a strongly correlated Quark-Gluon Plasma (sQGP), which subsequently undergoes a cross-over transition to hadronic matter with much lower energy densities. The sQGP flows like a nearly perfect fluid as measured by a very low value for the dimensionless ratio of the shear viscosity to entropy density in the fluid.  When the expansion rate of the system exceeds the scattering rate of the hadronic constituents, the matter free streams to the detectors. It is sometimes opined that the earliest stages of the nuclear collision are less likely to be understood than the latter stages. My view is that precisely the opposite may be true--one has some confidence that weak coupling techniques in QCD can be used to systematically compute features of early time dynamics. As the system evolves, weak coupling is less reliable and systematic computations can perhaps be performed for a limited sub-set of the system's properties\footnote{This does not mean we have all the answers for early time dynamics in weak coupling. An example is thermalization which is still poorly understood. However, it is less well understood in strong coupling frameworks.}.  Nevertheless, as we shall discuss later, a number of theoretical techniques, in combination with experimental results, can shed considerable light on many-body dynamics in the sQGP. 

In the following, we shall briefly summarize the theoretical ideas and the extant empirical evidence for each of the stages of the standard heavy ion model. It is impossible to do full justice to this topic in the limited space available for this talk and I will therefore have to frequently refer the reader to the available literature.  

\section{Before the collision: wee parton correlations in the Color Glass Condensate}

The appropriate asymptotics for multi-particle production in QCD is the Regge-Gribov asymptotics of the theory corresponding to $Q^2 = {\rm fixed}$, $x_{\rm Bj} \rightarrow 0$ and $s\rightarrow \infty$. In the framework of the theory where the parton model arises, this limit corresponds to very high parton phase space densities. In contrast, the more highly developed Bjorken asymptotics of the theory, $Q^2\rightarrow \infty$, $s\rightarrow \infty$ with $x_{\rm Bj} = {\rm fixed}$, corresponds to low occupation numbers and is more appropriate for the description of rare high momentum transfer processes. A natural consequence of Regge-Gribov asymptotics at small $x$ is parton saturation~\cite{GLR}, which, in the light front formulation of QCD, corresponds to maximal occupancy of $\sim 1/\alpha_S$ for gauge field modes below a momentum scale $Q_S$, the saturation scale. An equivalent description\cite{Iancu-Mueller} of parton saturation in the hadron rest frame corresponds to a small size (dipole) probe of size $\sim 2/Q_S$  scattering off a hotspot within the hadron with unit probability.

The high occupancy of saturated gluon modes (for $Q_S^2\gg \Lambda_{\rm QCD}^2$) suggests that the nuclear wavefunction at high energies (or small $x$) can be simply described by classical fields~\cite{MV}. Furthermore, because $Q_S$ is large, this high occupancy state is weakly coupled and its very non-perturbative behavior can be computed systematically order-by-order in $\alpha_S$ beyond the leading $1/\alpha_S$ classical contribution. The Wilsonian renormalization group (RG) equations that describe the evolution of this high occupancy state with energy are stochastic in nature, which explains why it is described to be a Color Glass Condensate (CGC)~\cite{reviews}. These RG equations, called JIMWLK equations~\cite{JIMWLK}, govern the energy evolution of n-point wee parton correlations in the hadron wavefunction. These carry much more information about QCD dynamics than parton distribution functions and are interesting to study in their own right.

\begin{figure}[t]
\centerline{
\includegraphics[height=6cm,width =5cm]{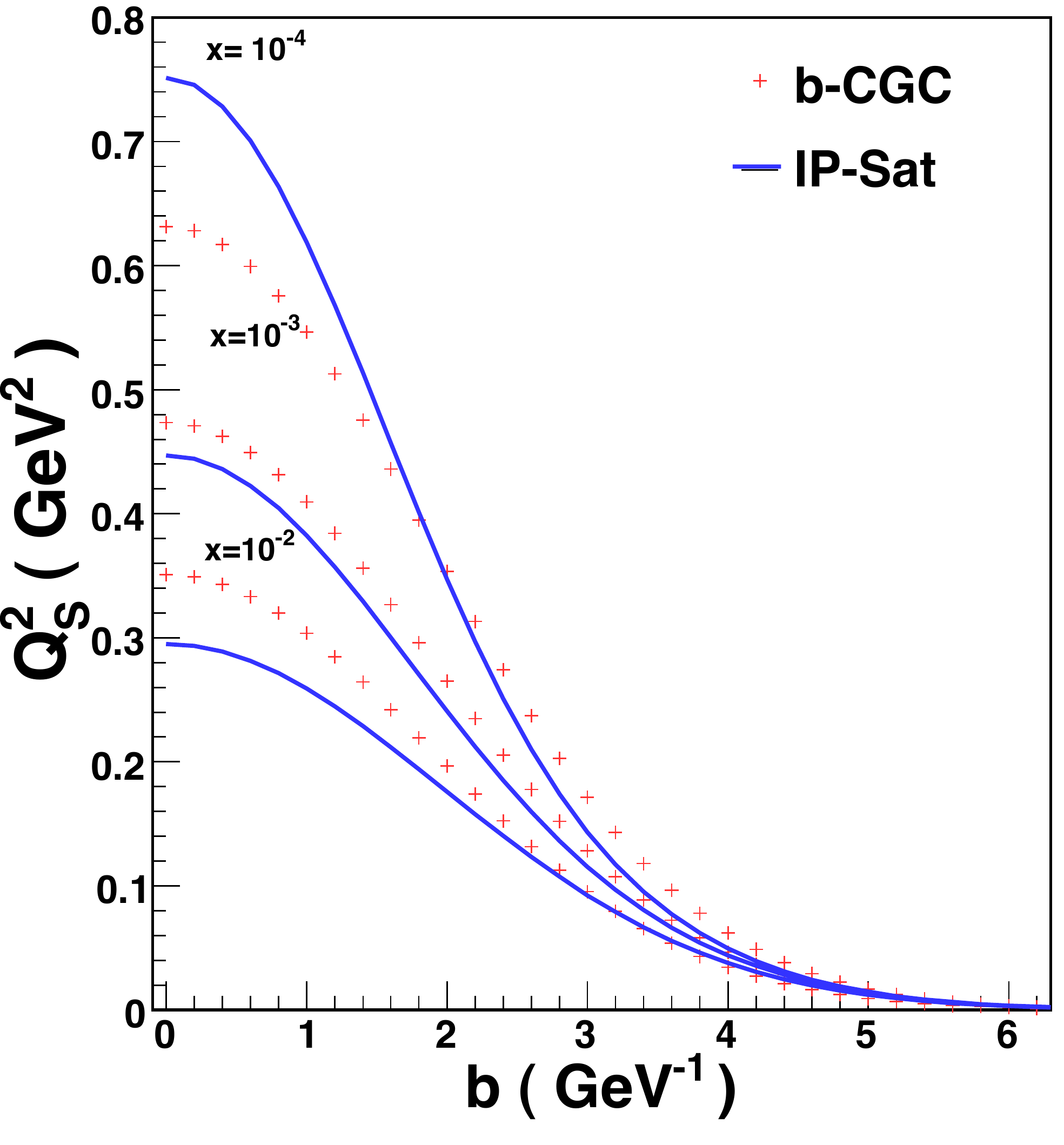}
\includegraphics[height=6cm,width =8cm]{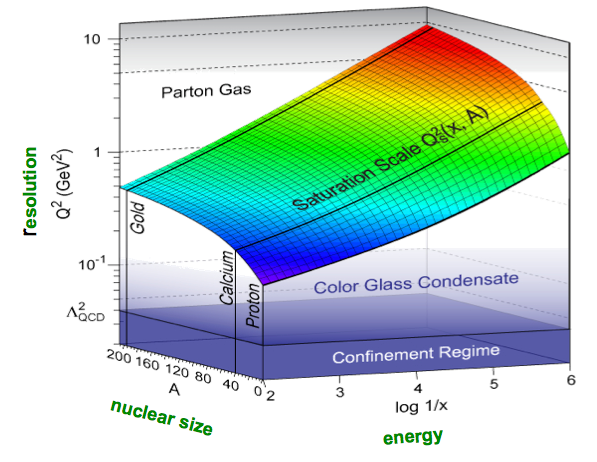}
}
\caption{Left: The saturation scale in the fundamental representation as a function of impact parameter extracted from HERA data in two different saturation models. Right: Picture of hadrons at high energies as resolved by a lepton probe.}
\label{fig:CGC}
\end{figure}

What is the evidence for parton saturation and the CGC ? An early remarkable phenomenon was geometrical scaling of the inclusive cross-section with $Q^2/Q_S^2$~\cite{Peschanski}. Phenomenological models that incorporate the physics of parton saturation do a very good job of describing HERA inclusive, diffractive and exclusive final states~\cite{Machado}, fixed target e+A inclusive data~\cite{KLV} and RHIC A+A multiplicities~\cite{Lappi-review} with an economical and consistent set of parameters. An early success of the saturation picture at RHIC was the description of the centrality dependence of the multiplicity distribution in A+A collisions~\cite{Kharzeev-Nardi}. The phenomenological saturation models have been shown to give a good description of the LHC p+p data including the n-particle multiplicity distribution\cite{Lev-Rez,Tribedy-RV}. 
Fig.~\ref{fig:CGC} (left) shows the saturation scale versus impact parameter for three different values of $x$ in p+p collisions extracted from HERA data in two different saturation models. The figure on the right shows the saturation scale as a function of $x$ and $A$. Both plots show the saturation scale in the fundamental representation. In the adjoint, relevant for hadronic collisions, one multiplies the value of $Q_S^2$ shown by $9/4$.

While the phenomenological models work very well, it is more satisfactory to confront next-to-leading order CGC predictions with the data. There has been quite a bit of progress in this direction. The Balitsky-Kovchegov (BK)~\cite{Balitsky-Kovchegov} equation, the large $N_c$ "mean field" equation in the CGC for the dipole forward scattering amplitude (corresponding to a 2-point Wilson line correlator) has now been computed at the next-to-leading log (NLL) level~\cite{Balitsky-Chirilli}. An evolution equation including the running coupling corrections in the NLL expressions~\cite{Albacete-Kovchegov} gives good agreement with the HERA inclusive data~\cite{Albacete-etal},  e+A inclusive data~\cite{Dusling-GLV} and the forward single inclusive deuteron+gold data from RHIC~\cite{Albacete-Marquet}. More interestingly, it describes~\cite{Albacete-Marquet2} the systematics of the disappearance of the away side hadron in di-pion production in deuteron-gold collisions at forward rapidities observed by the STAR collaboration~\cite{STAR-dA}. While theoretical caveats~\cite{Adrian-Jamal,Fabio-Bowen-Feng} to the predicted result may be quantitatively important, they are unlikely to alter its qualitative features. A more serious obstacle to this interpretation is from the possibility~\cite{Strikman-Vogelsang} that a azimuthal angle independent pedestal effect could mask this result. A clean experiment which will be able to settle the issue conclusively is to measure the same final state at a future electron-ion collider~\cite{Feng-Bowen}. 

In A+A collisions, most saturation models made predictions for the LHC that were on the low side by $\sim 20$\%~\cite{Armesto} compared to the recently released ALICE data~\cite{Alice-mult} for Pb+Pb collisions at 2.76 TeV/nucleon. Integrated multiplicities themselves are not terribly sensitive to dynamics. A 20\% discrepancy in the multiplicity corresponds to a corresponding discrepancy of 10\% in $Q_S$, which can easily be accounted for in revised fits. Indeed, when nuclear geometry effects are included, the NLO-BK model predicted a centrality dependence of the multiplicity~\cite{Albacete-Dumitru} that shows excellent agreement with the ALICE data.  

While each of these comparisons of CGC/saturation models to data in e+p/A, p+p, d+A and A+A collisions can be (and are) challenged, taken together they provide a remarkably consistent and detailed picture of a significant body of data suggesting that saturation effects have been seen and are essential to understanding multi-particle production at high energy colliders.  At the LHC (and in forward observables at RHIC), the saturation scale is large enough that weak coupling techniques are applicable. This therefore suggests the exciting possibility that physics previously thought inaccessible is now amenable to systematic computation. 

\section{The little bang and the Glasma}

It was understood early on in QCD that multi-particle production in hadron-hadron collisions arises from the scattering of their wee parton clouds~\cite{Bj-DESY}. In the CGC, these are classical fields with their field strengths localized on the Lorentz contracted widths of the nuclei, so to lowest order in the coupling constant (of order $1/\alpha_S$), the scattering of two CGCs is described by the collision of two classical fields. This is equivalent to solving Yang-Mills equations with initial conditions determined by the classical fields for each nucleus~\cite{KMW}. These classical equations have been solved and the single inclusive~\cite{KV} and more recently the double inclusive~\cite{StasLV} distributions determined. To this lowest order, the classical equations (and particle distributions) are boost invariant. The gauge field configurations are color screened on distances of order $1/Q_S$ and right after the collisions are dominated by longitudinal color-electric and color-magnetic fields that carrry Chern-Simons charge~\cite{Kharzeev-KV}. The QCD matter corresponding to early time non-equilibrium dynamics has been dubbed "the Glasma"~\cite{Lappi-M}. 

The classical configurations in the Glasma are far from the whole story. Quantum fluctuations profoundly enrich this picture\footnote{This is fortuitous because the naive leading order picture is not what is seen in experiments.}. Before the collision, quantum fluctuations generate large logarithms $\alpha_S\ln(1/x_{1,2})$ that are of the same order as the leading classical term and therefore have to be resummed at each order in perturbation theory. At each loop order, there are also large multiple scattering contributions that need to be summed to all orders in $\alpha_S n$, where $n\sim 1/\alpha_S$ is the density of parton sources. High energy factorization theorems proven recently for inclusive quantities show that the leading logs in $x$ and multiple scattering contributions can be factorized\footnote{The underlying basis for this factorization is the observation that the computation of inclusive quantities in field theories with strong time dependent sources can be formulated as an initial value problem~\cite{Gelis-RV}.} from the final state dynamics into universal density matrices that satisfy the JIMWLK equation~\cite{Gelis-Lappi-RV} convoluted with the inclusive observable computed at leading order. Thus correlators of Wilson lines extracted from e+A scattering can in principle be used as inputs into computing A+A final states. 

The high energy factorization theorems are a powerful tool to compute the early stage space-time evolution of  quantities such as the stress-energy tensor or correlators of the stress-energy tensor {\it ab initio} in heavy ion collisions. They allow one to compute for instance long range rapidity correlations of inclusive observables that are sensitive to the earliest times in the collision. A consequence of the formalism is the Glasma flux tube picture~\cite{Dumitru-GMV} where n-particle correlations are simply proportional to  $(S_\perp/(1/Q_S^2))^{1-n}$ for $n\geq 2$, the ratio of the nuclear transverse area $S_\perp$ to the area of a flux tube $1/Q_S^2$. These geometrical correlations give rise~\cite{Gelis-LM} to the negative binomial distribution which explains multiplicity distributions~\cite{Tribedy-RV} and forward-backward multiplicity correlations~\cite{Armesto-MP} in hadronic collisions. 

\begin{figure}[t]
\centerline{
\includegraphics[height=4cm,width =7cm]{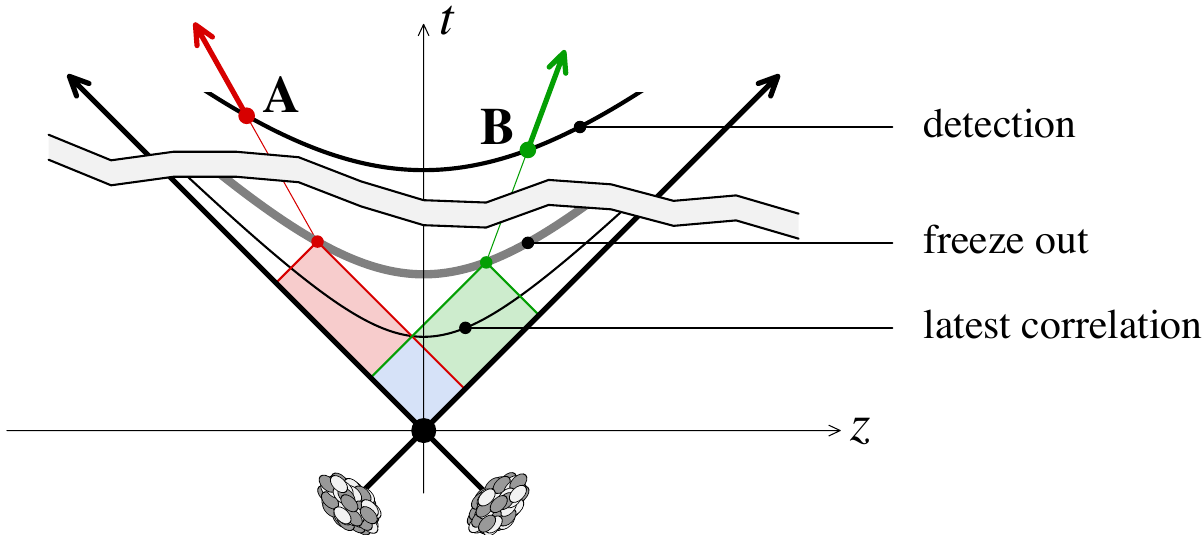}
\includegraphics[height=4cm,width =7cm]{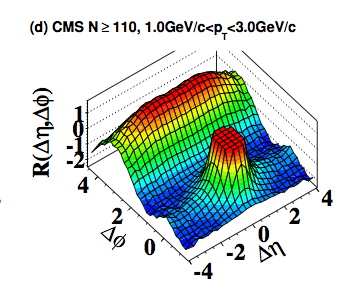}
}
\caption{Left: Plot illustrates how long range rapidity correlations are causally limited to provide information on 
early time dynamics. Right: A collimated structure around $\Delta \phi\approx 0$ that is long range in rapidity seen by the CMS collaboration in high multiplicity p+p collisions at c.m.s energies of 7 TeV. This ridge-like structure is not seen in Monte-Carlo models while the away side peak at $\Delta\phi \approx\pi$ is well reproduced in these models.}
\label{fig:Glasma}
\end{figure}

Long range rapidity correlations that are  sharply collimated around $\Delta \Phi\approx 0$ were 
observed at RHIC~\cite{RHIC-ridge}. These structures, called "ridges" based on their visual appearance, are present both for triggered and untriggered two particle correlations. These ridges are not present in peripheral events and are most prominent in the most central events. Long range rapidity correlations are interesting because they are a "chronometer" of strong color field dynamics at early times in hadronic collisions. This is illustrated in fig.~\ref{fig:Glasma} (left).

In the Glasma flux tube picture~\cite{Dumitru-GMV}, the ridge phenomenon is a consequence of the long range rapidity correlations in flux tubes of transverse size $1/Q_S$ formed early in the collision that are subsequently boosted in the final state by radial flow~\cite{Voloshin-Shuryak}. The radial flow provides the near side angular collimation. The combination of flux tube structures in the initial state and radial flow gives an excellent description of the published RHIC data~\cite{Gavin-Moschelli,Werner-Kodama}. The ALICE collaboration has presented preliminary evidence of a ridge in Pb+Pb collisions~\cite{Schukraft}--we would expect the amplitude to follow the trend seen at RHIC and be larger than the RHIC values. At the LHC, one expects the ridge correlations  to show deviations from boost invariance~\cite{Dusling-GLV} for $\Delta \eta > 4$ units and the amplitude of the ridge to decrease with increasing transvere momentum for the pairs~\cite{Raju-QMtalk}.

In a recent paper, the CMS collaboration announced the discovery of a ridge in high multiplicity events in p+p collisions~\cite{CMS-ridge} shown in fig.~\ref{fig:Glasma} (right). This effect was predicted~\cite{Dumitru} based on our formalism~\cite{Dusling-GLV} but was not advertised because it was a small effect. We  subsequently showed the Glasma flux tubes predictions are in qualitative~\cite{Dumitru-etal} and quantitative agreement with the experimental data~\cite{Dusling-RV}. We note that long range 
forward-backward correlations were predicted for p+p collisions at the LHC in a related framework~\cite{Pajares}.

Another effect that is very sensitive to properties of the Glasma is the Chiral Magnetic Effect (CME)~\cite{Kharzeev-etal}. This is clever idea that suggests that sphaleron transitions in deconfined matter subject to an external magnetic field can lead to charge separation in the direction of the magnetic field that locally breaks P and CP though of course it is preserved in the event as a whole. The STAR collaboration published data on same and opposite charge separations which appeared consistent with the CME~\cite{STAR-CP}. Several recent papers however suggest alternative analyses of the STAR data~\cite{CME-alternate}--the RHIC low energy run may help clarify which interpretation is correct. Nevertheless, the CME has stirred a lot of theoretical interest and has ramifications for fields outside heavy ion collisions~\cite{Fraga}. 

How the Glasma thermalizes to a QGP is an outstanding problem that has not been solved even in weak coupling despite much theoretical progress in recent years. An important ingredient is the role of instabilities~\cite{Mrowczynski}. In the CGC framework, unstable quantum fluctuations play an essential role in close analogy to the situation in inflationary cosmology~\cite{cosmo-reviews}. Rapidly growing quantum fluctuations can be resummed; in a toy scalar theory, the corresponding energy-momentum tensor obeys ideal relativistic hydrodynamics~\cite{Dusling-EGV}. If these considerations can be extended to a gauge theory, one could have "hydrodynamic-like" flow without early thermalization. 

\section{The perfect fluid}

A key result of the RHIC experiments is the large flow measured suggesting that hydrodynamics is applicable. Hydrodynamics is the right effective field theory to describe the long wavelength, late time behavior of quantum field theory. The surprise at RHIC is that it  works at much earlier times than simple estimates would suggest. A powerful measure of the degree of flow is $v_2$, the second moment of the anisotropy in the single particle distribution. Specifically, it measures how efficiently hot matter converts spatial anisotropies in the initial distribution to momentum anisotropies at later times--the most efficient way to do this is by applying ideal relativistic hydrodynamics. 

A quantitative measure of the efficiency of flow is given by the ratio $\eta/s$ of the shear viscosity to the entropy density in the fluid. For simplified boost-invariant ("Bjorken") hydrodynamics, the viscous contribution to the evolution in proper time $\tau$  of the energy density relative to the ideal term is given by $\eta/s /\tau T$; because $1/\tau T \sim O(1)$, this contribution is small when $\eta/s<< 1$. In kinetic theory, $\eta/s \approx \tau_{\rm relax.}/\tau_{\rm quant.}$, in units of $\hbar/k_B$, where $\tau_{\rm relax.}$ is the momentum relaxation time in a fluid in response to a shear stress and $\tau_{\rm quant.}$ is the thermal Compton wavelength in the fluid. One therefore expects naively a lower bound of unity on $\eta/s$. However, this expectation is not correct because there is no reason kinetic theory is relevant in this domain. In fact, for strongy correlated systems one expects kinetic theory to fail\footnote{Albeit, one should note that kinetic theory often does much better (as is the case for cold atomic gases) than its condition of applicability would suggest.}. Nevertheless, on very general grounds one expects a lower bound on $\eta/s$. Is there such a bound and how does the value extracted from RHIC experiments compare ?

While the answer to the first part of the question is not known definitively, there is a conjectured lower bound\footnote{There are recent suggestions that in certain higher derivative gravity theories an even lower bound may be attained~\cite{Myers}.} of $\eta/s= 1/4\pi$~\cite{Son-etal}. This bound was derived for N=4 supersymmetric Yang-Mills (SYM) theory and makes use of the Maldacena AdS/CFT conjecture relating SYM field theory in 4 dimensions to 10 dimensional classical gravity in a background of D3 branes~\cite{Maldacena}. Specifically, the viscosity in SYM is related to the absorption cross-section of a graviton on a black brane, which like the entropy density, is proportional to the area of the brane.

\begin{figure}[t]
\centerline{
\includegraphics[height=5cm,width =7cm]{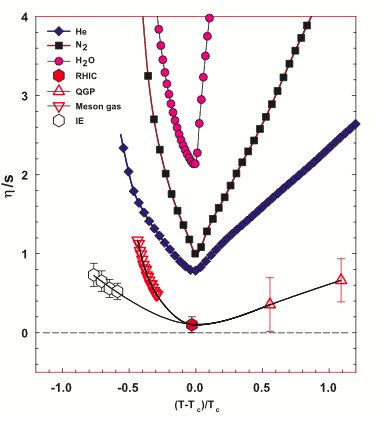}
\includegraphics[height=5cm,width =7cm]{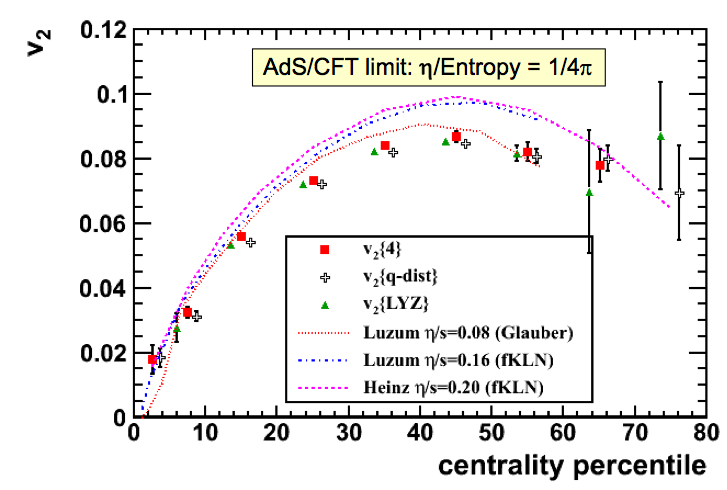}
}
\caption{Left: Behavior of $\eta/s$ in a number of fluids measured relative to deviation from the critical 
temperature~\cite{Lacey}. Right: Data on $v_2$ from the ALICE collaboration~\cite{Alice-v2talk} compared to 
predictions from viscous hydrodynamic models.}
\label{fig:PerfectFluid}
\end{figure}

What is $\eta/s$ extracted empirically ? For this one relies on viscous hydrodynamic models--the algorithms for these have developed significantly~\cite{visc-hydro-reviews}. While there are uncertainties related to the initial conditions and other details of the numerical implementation, its probably safe to say that $\eta/s \leq 0.4$~\cite{Schafer-Teaney}. A comparison of $\eta/s$ in various fluids at temperatures close to the critical temperature is shown in fig.~\ref{fig:PerfectFluid} (left). These values of $\eta/s$ are much lower than that of water or even liquid helium but are comparable to those of strongly correlated lithium-6 atoms in the unitarity limit. It is remarkable that two systems whose temperatures differ by $10^{19}$ and viscosities by $10^{26}$ flow nearly identically. 

Most weak coupling estimates of $\eta/s$ in the QGP give values for $\eta/s$ that are significantly larger~\cite{Moore}. Does this rule out a weak coupling description of flow completely? One way out is if the system had a small ``anomalous" viscosity (as seen in electromagnetic plasmas) which mimics a small kinematic shear viscosity~\cite{Asakawa-BM}.  If, as discussed previously, quantum corrections to the Glasma lead to ideal hydrodynamic behavior, this could provide a quantitative mechanism for the "anomalous" early time viscosity. Detailed studies are feasible in near future on how much the empirical lower bound on $\eta/s$ can be moved upwards. 

The ALICE collaboration has released first results on $v_2$ seen in Pb+Pb collisions at $\sqrt{s} = 2.76$ TeV/nucleon~\cite{ALICE-v2}. The integrated $v_2$ is about 30\% higher than at RHIC but $v_2(p_\perp)$ is nearly identical to that seen at RHIC. This is what one would expect if a nearly perfect fluid had been formed at RHIC and is consistent with the predictions from viscous hydro models~\cite{Luzum}--see fig.~\ref{fig:PerfectFluid} (right). If significant entropy were generated, it should impact the centrality dependence of the multiplicity distribution, which is consistent with initial state predictions for {\it both} RHIC and the LHC. This centrality dependence also has the potential of constraining the thermalization time of the Glasma~\cite{Dumitru-MN}. Interesting measurements at RHIC (and in near future at LHC) on $v_2$ fluctuations and higher moments of the anisotropy distribution are quite sensitive to fine details of the initial conditions and final state effects~\cite{Sorensen}. For instance, the ratio $v_4/v_2^2$ gives very different results for different {\it ans\"{a}tze} of the viscous corrections to the single particle distributions at freeze-out~\cite{Ollitrault-talk}.

\section{Hard probes of QCD matter}

Hydrodynamics paints broad brush strokes of the dynamics of strongly correlated matter. To obtain further insight, one looks to hard probes of the QCD medium. Understanding the dynamics of the medium from these is very interesting albeit very challenging as well because a colored hard probe is typically sensitive to the entire time history of the system. At RHIC, it was observed that 
$R_{AA}$--the ratio (normalized by the number of binary nucleon-nucleon collisions) of the inclusive hadron spectrum in A+A collisions relative to the same quantity in p+p collisions--was suppressed by a factor of 5 for $\pi^0$'s and is nearly flat out to the highest $p_\perp$'s 
measured at RHIC~\cite{jetquenching}. A control deuteron+gold measurement~\cite{dA} established that this phenomenon was a final state effect arising from the interaction of partons with the colored medium. 

"Jet quenching" was suggested as a probe of the QGP a long time ago~\cite{Bj-Appel} and radiative energy loss by partons traversing the medium was identified as a weak coupling mechanism that would explain it~\cite{Gyulassy-Wang,BDMPS}. The energy loss of partons can be related to a transport coefficient $\hat{q}$ that characterizes the properties of the medium. There are several formalisms based on the original works that have been developed to confront the heavy ion data on jet quenching~\cite{quenching-reviews}. These differ quite a bit in detail and give values of $\hat{q}$ that differ considerably--for a clear and interesting recent discussion of the sources of some of these differences, see Ref.~\cite{Gale-Huot}.

The flat behavior of $R_{AA}$, the large values of $v_2(p_\perp)$ out to large $p_\perp$,  and the 
unexpectedly large energy loss of heavy quarks are difficult to reconcile in a weak coupling framework without fine tuning. An example of a fine tuning argument is one that explains the flat behavior of $R_{AA}$ at RHIC as arising from a combination of effects, in particular phase space constraints at large $x_T$, that modify the weak coupling prediction that the suppression goes away with increasing $p_\perp$. This should be different at the LHC where there is no large $x_T$ constraint. The first $R_{AA}$ data at LHC for charged hadrons indeed appear to favor this interpretation~\cite{ALICE-jetquenching}. One would like however to see the spectrum of $\pi^0$'s and a wider range in $p_\perp$ before more definitive statements are made. An important factor that should be kept in mind while interpreting $R_{AA}$ is the very large initial state  gluon shadowing at LHC energies~\cite{Paukkunen,Albacete-Marquet}. 

The apparent strong opacity of the QGP has triggered a considerable amount of work~\cite{Iancu-review} to explain light and heavy quark energy loss in the same AdS/CFT framework that gave us the lower bound on $\eta/s$. There are two sorts of treatments here. In one, the jet production mechanism is treated perturbatively and  factorized from the non-perturbative interaction of the jet with the medium~\cite{Liu-etal}. The latter is expressed in terms of correlators of Wilson lines which is computed using the AdS/CFT correspondence. In the other, jet quenching is treated completely in the AdS/CFT framework, typically by looking at the energy loss of heavy quarks~\cite{Yaffe-Gubser-Teaney}. QCD is however significantly different from N=4 SYM, especially in the region around $T_c$, where one notices strong conformal symmetry breaking effects. Until one has a good candidate for a QCD dual in the AdS/CFT framework, perhaps its value is primarily to provide guidance into dynamical questions that one cannot easily answer in the QCD plasma in the region where weak coupling is suspect. For other approaches in addressing this question, see for instance ref.~\cite{Pisarski}. 

\begin{figure}[t]
\centerline{
\includegraphics[height=5cm,width =7cm]{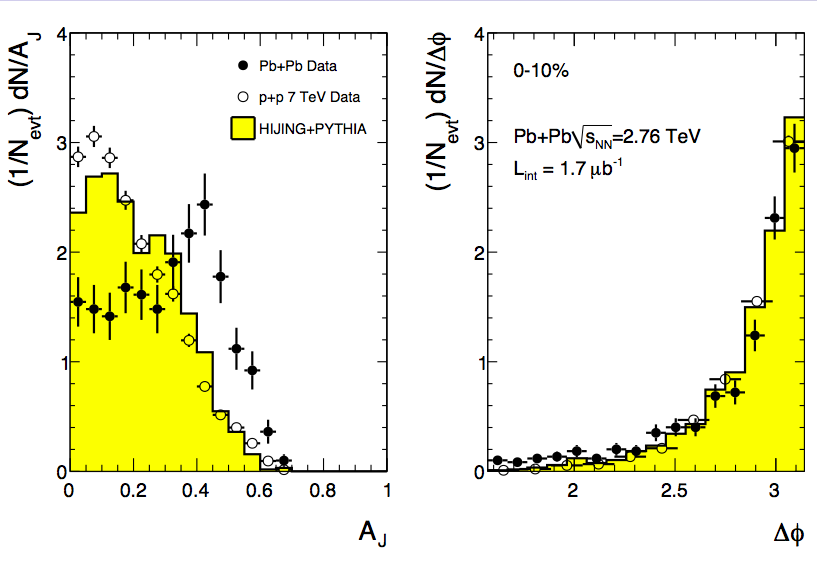}
\includegraphics[height=5cm,width =7cm]{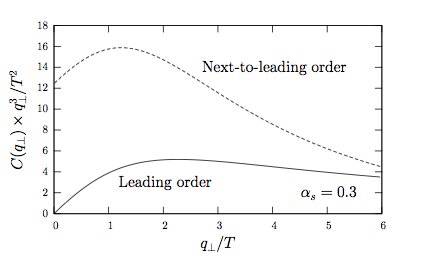}
}
\caption{Left: Data from the ATLAS collaboration showing significant jet asymmetry for central Pb+Pb collisions. 
The data also shows that despite the asymmetry, the jets are back-to-back. Right: Next-to-leading order estimate compared to leading order estimate of contributions to the second moment ($\hat{q}$) of the elastic scattering rate of a hard jet with a hot medium for transverse momenta close to the medium temperature.}
\label{fig:Jets}
\end{figure}

It is important to understand fully the limits of the applicability of the weak coupling framework. For instance, present treatments of the parton interaction with the medium are at tree level. An interesting quantity to compute is the elastic scattering rate for a hard particle in the medium; the transport coefficient $\hat{q}$ is the second moment of this quantity. The first NLO $O(g)$ computation of this quantity has been reported recently~\cite{Caron-huot}; the correction to the leading order result is significant. The estimate is shown in fig.~\ref{fig:Jets} (right). This raises the question whether resummations {\it a la} the pressure in finite temperature field theory are required. 

As promised, jets have been produced in copious amounts in Pb+Pb collisions at the LHC.  The ATLAS and CMS collaborations confirm observation of jet quenching in di-jets~\cite{ATLAS-jet,CMS-jet}. Looking at trigger jets with $E_{T1} > 100$ GeV and away side jets in the opposite hemisphere with $E_{T2} > 25$ GeV as a function of the variable $A_J = (E_{T1}-E_{T2})/E_{T1}+E_{T2}$, they find a centrality dependent di-jet asymmetry that is strongest for the most central collisions. At the same time, the vast majority of the di-jets are back-to-back--see fig.~\ref{fig:Jets} (left). A simple explanation~\cite{Solana-MW} that makes intuitive sense follows from the observation that firstly, the softest components of the jet shower decohere sooner and secondly, are more easily deflected out of the jet cone by scattering off the hot medium thereby degrading the energy of the jet, while preserving its angular structure. The first estimates of $\hat{q}$ are consistent with a weak coupling scenario. Clearly, a lot more data can be anticipated in the very near future that will add considerable depth to the existing picture of jet quenching.

\section{Looking ahead}

I have outlined here the bare bones of a standard model of heavy ion collisions. Even within this structure, there is much that we don't understand, but we have a better idea of what that is. With LHC very quickly showing us what is possible and the high luminosity and detector upgrades at RHIC on track, we can be optimistic the outstanding issues will be solved. Because many-body QCD is a rich and subtle science, a natural extension of on-going research on hot matter is to explore the complementary many body dynamics of "cold" nuclear matter with a high energy, high luminosity electron-ion collider.


\begin{thebibliography}{99}
\bibitem{Wilczek} F.~Wilczek,
  %``What QCD tells us about nature - and why we should listen,''
  Nucl.\ Phys.\  A {\bf 663}, 3 (2000)
  [arXiv:hep-ph/9907340].

\bibitem{Cabibo-Parisi}N.~Cabibbo and G.~Parisi,
  %``Exponential Hadronic Spectrum And Quark Liberation,''
  Phys.\ Lett.\  B {\bf 59}, 67 (1975).

\bibitem{Collins-Perry}J.~C.~Collins and M.~J.~Perry,
  %``Superdense Matter: Neutrons Or Asymptotically Free Quarks?,''
  Phys.\ Rev.\ Lett.\  {\bf 34}, 1353 (1975).

\bibitem{Shuryak}E.~V.~Shuryak,
  %``Quark-Gluon Plasma And Hadronic Production Of Leptons, Photons And
  %Psions,''
  Phys.\ Lett.\  B {\bf 78}, 150 (1978).

\bibitem{Philipsen}O. Philipsen, talk at this conference.

\bibitem{GLR}L.V. Gribov, E.M. Levin, M.G. Ryskin, Phys. Rept. {\bf 100}, 1 (1983); A.H. Mueller, J-W. Qiu, Nucl. Phys. {\bf B} {\bf 268}, 427 (1986).

\bibitem{Iancu-Mueller}E.~Iancu and A.~H.~Mueller,
  %``From color glass to color dipoles in high-energy onium onium  scattering,''
  Nucl.\ Phys.\  A {\bf 730}, 460 (2004).

\bibitem{MV}L.D. McLerran, R. Venugopalan,  Phys. Rev. {\bf D} {\bf 49}, 2233 (1994); {\it ibid.} {\bf 49}, 3352 (1994); {\it ibid.} {\bf 50}, 2225 (1994).

\bibitem{reviews}E. Iancu, R. Venugopalan, Quark Gluon Plasma 3, Eds. R.C. Hwa, X.N. Wang,
  World Scientific, hep-ph/0303204; H.~Weigert,
  %``Evolution at small x(bj): The color glass condensate,''
  Prog.\ Part.\ Nucl.\ Phys.\  {\bf 55}, 461 (2005); F. Gelis, E. Iancu, J. Jalilian-Marian, R. Venugopalan, arXiv:1002.0333.

\bibitem{JIMWLK}Jalilian-Marian, A. Kovner, A. Leonidov, H. Weigert, Nucl. Phys. {\bf B}
  {\bf 504}, 415 (1997); {\it ibid.}, Phys. Rev. {\bf D}
  {\bf 59}, 014014 (1999); E. Iancu, A. Leonidov, L.D. McLerran, Nucl. Phys. {\bf A} {\bf 692}, 583
  (2001); E. Ferreiro, E. Iancu, A. Leonidov, L.D. McLerran, Nucl. Phys. {\bf A} {\bf
  703}, 489 (2002).

\bibitem{Peschanski}R. Peschanski, talk at this conference.

\bibitem{Machado}M. V. Machado, talk at this conference, arXiv:1011.5190 [hep-ph]; H.~Kowalski, L.~Motyka, G.~Watt,
  %``Exclusive diffractive processes at HERA within the dipole picture,''
  Phys.\ Rev.\  D {\bf 74}, 074016 (2006).

\bibitem{KLV}H.~Kowalski, T.~Lappi, R.~Venugopalan,
  %``Nuclear enhancement of universal dynamics of high parton densities,''
  Phys.\ Rev.\ Lett.\  {\bf 100}, 022303 (2008).

\bibitem{Lappi-review}T.~Lappi,
  %``Small x physics and RHIC data,''
  arXiv:1003.1852 [hep-ph].

\bibitem{Kharzeev-Nardi}D.~Kharzeev and M.~Nardi,
  %``Hadron production in nuclear collisions at RHIC and high density QCD,''
  Phys.\ Lett.\  B {\bf 507}, 121 (2001).

\bibitem{Lev-Rez}E.~Levin, A.~H.~Rezaeian,
  %``Gluon saturation and inclusive hadron production at LHC,''
  Phys.\ Rev.\  D {\bf 82}, 014022 (2010).

\bibitem{Tribedy-RV}P.~Tribedy and R.~Venugopalan,
  %``Saturation models of HERA DIS data and inclusive hadron distributions in
  %p+p collisions at the LHC,''
  arXiv:1011.1895 [hep-ph].

\bibitem{Balitsky-Kovchegov}I. Balitsky, Nucl. Phys. {\bf B} {\bf 463}, 99 (1996); Yu.V. Kovchegov, Phys. Rev. {\bf D} {\bf 61}, 074018 (2000).

\bibitem{Balitsky-Chirilli}I.~Balitsky,  G.~A.~Chirilli,
  %``Next-To-Leading Order Evolution Of Color Dipoles,''
  Phys.\ Rev.\  D {\bf 77}, 014019 (2008); Y.~V.~Kovchegov, H.~Weigert,
  %``Triumvirate of running couplings in small-x evolution,''
  Nucl.\ Phys.\  A {\bf 784}, 188 (2007); G.~Beuf,
  %``Universal behavior of the gluon saturation scale at high energy including
  %full NLL BFKL effects,''
  arXiv:1008.0498 [hep-ph].

\bibitem{Albacete-Kovchegov}J.~L.~Albacete, Y.~V.~Kovchegov,
  %``Solving High Energy Evolution Equation Including Running Coupling
  %Corrections,''
  Phys.\ Rev.\  D {\bf 75}, 125021 (2007).

\bibitem{Albacete-etal}J.~L.~Albacete, N.~Armesto, J.~G.~Milhano , C.~A.~Salgado,
  %``Non-linear QCD meets data: A global analysis of lepton-proton scattering
  %with running coupling BK evolution,''
  Phys.\ Rev.\  D {\bf 80}, 034031 (2009).


\bibitem{Albacete-Marquet}J.~L.~Albacete, C.~Marquet,
  %``Single Inclusive Hadron Production at RHIC and the LHC from the Color Glass
  %Condensate,''
  Phys.\ Lett.\  B {\bf 687}, 174 (2010).

\bibitem{Albacete-Marquet2}J.~L.~Albacete,  C.~Marquet,
  %``Azimuthal correlations of forward di-hadrons in d+Au collisions at RHIC in
  %the Color Glass Condensate,''
  Phys.\ Rev.\ Lett.\  {\bf 105}, 162301 (2010); see also, K.~Tuchin,
  %``Rapidity and centrality dependence of azimuthal correlations in
  %Deuteron-Gold collisions at RHIC,''
  Nucl.\ Phys.\  A {\bf 846}, 83 (2010).

\bibitem{STAR-dA}
  E.~Braidot  [STAR Collaboration],
  %``Suppression of Forward Pion Correlations in d+Au Interactions at STAR,''
  arXiv:1005.2378 [hep-ph].

\bibitem{Adrian-Jamal}A.~Dumitru and J.~Jalilian-Marian,
  %``Two-particle correlations in high energy collisions and the gluon
  %four-point function,''
  Phys.\ Rev.\  D {\bf 81}, 094015 (2010).

\bibitem{Fabio-Bowen-Feng}F.~Dominguez, B.~W.~Xiao and F.~Yuan,
  %``kt-factorization for Hard Processes in Nuclei,''
  arXiv:1009.2141 [hep-ph].

\bibitem{Strikman-Vogelsang}M.~Strikman and W.~Vogelsang,
  %``Multiple parton interactions and forward double pion production in pp and
  %dA scattering,''
  arXiv:1009.6123 [hep-ph].

\bibitem{Feng-Bowen}B.~W.~Xiao and F.~Yuan,
  %``Non-Universality of Transverse Momentum Dependent Parton Distributions at
  %Small-x,''
  Phys.\ Rev.\ Lett.\  {\bf 105}, 062001 (2010).

\bibitem{Armesto}N.~Armesto {\it et al.},
  %``Heavy Ion Collisions at the LHC - Last Call for Predictions,''
  J.\ Phys.\ G {\bf 35}, 054001 (2008).

\bibitem{Alice-mult}ALICE Collaboration,
  %``Centrality dependence of the charged-particle multiplicity density at
  %mid-rapidity in Pb-Pb collisions at sqrt(sNN) = 2.76 TeV,''
  arXiv:1012.1657 [nucl-ex].

\bibitem{Albacete-Dumitru}J.~L.~Albacete and A.~Dumitru,
  %``A model for gluon production in heavy-ion collisions at the LHC with rcBK
  %unintegrated gluon densities,''
  arXiv:1011.5161 [hep-ph].

\bibitem{Bj-DESY}J.~D.~Bjorken,
  %``Hadron Final States In Deep Inelastic Processes,''
  Lect.\ Notes Phys.\  {\bf 56}, 93 (1976).

\bibitem{KMW}A.~Kovner, L.~D.~McLerran and H.~Weigert,
  %``Gluon production at high transverse momentum in the McLerran-Venugopalan
  %model of nuclear structure functions,''
  Phys.\ Rev.\  D {\bf 52}, 3809 (1995).

\bibitem{KV}A.~Krasnitz and R.~Venugopalan,
  %``Non-perturbative computation of gluon mini-jet production in nuclear
  %collisions at very high energies,''
  Nucl.\ Phys.\  B {\bf 557}, 237 (1999).

\bibitem{StasLV}T. Lappi, S. Srednyak, R. Venugopalan, JHEP {\bf 1001} 066 (2010).

\bibitem{Kharzeev-KV}D. Kharzeev, A. Krasnitz, R. Venugopalan, Phys. Lett. {\bf B} {\bf 545}, 298
  (2002).

\bibitem{Lappi-M}T. Lappi, L.D. McLerran, Nucl. Phys. {\bf A} {\bf 772}, 200 (2006).

\bibitem{Gelis-RV}F.~Gelis,  R.~Venugopalan,
  %``Particle production in field theories coupled to strong external
  %sources,''
  Nucl.\ Phys.\  A {\bf 776}, 135 (2006); {\it ibid.} A {\bf 779}, 177 (2006).

\bibitem{Gelis-Lappi-RV}F. Gelis, T. Lappi, R. Venugopalan, Phys. Rev. {\bf D} {\bf 78}, 054019
  (2008); {\it ibid.},   {\bf 78}, 054020 (2008); {\it ibid.}, {\bf 79}, 094017 (2009).

\bibitem{Dumitru-GMV}A. Dumitru, F. Gelis, L. McLerran, R. Venugopalan, Nucl. Phys. {\bf A} {\bf
  810}, 91 (2008).

\bibitem{Gelis-LM}F. Gelis, T. Lappi,  L. McLerran, Nucl. Phys. A828 (2009) 149.

\bibitem{Armesto-MP}N.~Armesto, L.~McLerran, C.~Pajares,
  %``Long range forward-backward correlations and the color glass condensate,''
  Nucl.\ Phys.\  A {\bf 781}, 201 (2007); T.~Lappi,  L.~McLerran,
  %``Long range rapidity correlations as seen in the STAR experiment,''
  Nucl.\ Phys.\  A {\bf 832}, 330 (2010).

\bibitem{RHIC-ridge}B.~I.~Abelev {\it et al.}  [STAR Collaboration],
  %``Long range rapidity correlations and jet production in high energy nuclear
  %collisions,''
  Phys.\ Rev.\  C {\bf 80}, 064912 (2009); \bibitem{Adamsa5}
{J. Adams, {\it et al.}}, [STAR Collaboration] Phys. Rev. {\bf C} {\bf 73}, 064907
  (2006); A.~Adare {\it et al.}  [PHENIX Collaboration],
  %``Dihadron azimuthal correlations in Au+Au collisions at sqrt(s_NN)=200
  %GeV,''
  Phys.\ Rev.\  C {\bf 78}, 014901 (2008);  B.~Alver {\it et al.}  [PHOBOS Collaboration],
  %``High transverse momentum triggered correlations over a large pseudorapidity
  %acceptance in Au+Au collisions at sqrt(s_NN)=200 GeV,''
  Phys.\ Rev.\ Lett.\  {\bf 104}, 062301 (2010).

\bibitem{Voloshin-Shuryak}S.A. Voloshin, Phys. Lett. {\bf B} {\bf 632}, 490 (2006); 
E.V. Shuryak, Phys. Rev. {\bf C} {\bf 76}, 047901 (2007); C.A. Pruneau, S. Gavin, S.A. Voloshin, Nucl. Phys. {\bf A} {\bf 802}, 107
  (2008).

\bibitem{Gavin-Moschelli}S. Gavin, L. McLerran, G. Moschelli, Phys. Rev. {\bf C} {\bf 79}, 051902
  (2009); G.~Moschelli and S.~Gavin,
%``Soft Contribution to the Hard Ridge in Relativistic Nuclear Collisions,''
Nucl.\ Phys.\  A {\bf 836}, 43 (2010).

\bibitem{Werner-Kodama}K.~Werner, I.~Karpenko, T.~Pierog, M.~Bleicher and K.~Mikhailov,
%``Event-by-Event Simulation of the Three-Dimensional Hydrodynamic Evolution
%from Flux Tube Initial Conditions in Ultrarelativistic Heavy Ion
%Collisions,''
arXiv:1004.0805 [nucl-th]; J.~Takahashi {\it et al.},
%``Topology studies of hydrodynamics using two particle correlation
%analysis,''
Phys.\ Rev.\ Lett.\  {\bf 103}, 242301 (2009).

\bibitem{Schukraft}J. Schukraft, talk at CERN Heavy Ion Seminar, December 2nd, (2010).

\bibitem{Dusling-GLV}K. Dusling, F. Gelis, T. Lappi, R. Venugopalan, Nucl. Phys. {\bf A} {\bf 836}, 159 (2010).

\bibitem{Dumitru}A.~Dumitru, in RIKEN-BNL Center Workshop on ``Progress in High pT
Physics at RHIC'', March 17 -- 19, 2010, RBRC Vol.\ 95, page 129.

\bibitem{Dumitru-etal}A.~Dumitru, K.~Dusling, F.~Gelis, J.~Jalilian-Marian, T.~Lappi, R.~Venugopalan,
  %``The ridge in proton-proton collisions at the LHC,''
  arXiv:1009.5295 [hep-ph].

\bibitem{Dusling-RV}K. Dusling and R. Venugopalan, in preparation.

\bibitem{Raju-QMtalk}F.~Gelis, T.~Lappi and R.~Venugopalan,
  %``Long range rapidity correlations and the ridge in A+A collisions,''
  Nucl.\ Phys.\  A {\bf 830}, 591C (2009)
  [arXiv:0907.4381 [hep-ph]].

\bibitem{CMS-ridge}CMS Collaboration, arXiv:1009.4122.

\bibitem{Pajares}P.~Brogueira, J.~Dias de Deus and C.~Pajares,
  %``Long range forward-backward rapiditiy correlations in proton-proton
  %collisions at LHC,''
  Phys.\ Lett.\  B {\bf 675}, 308 (2009).

\bibitem{Kharzeev-etal}D.~E.~Kharzeev, L.~D.~McLerran and H.~J.~Warringa,
  %``The effects of topological charge change in heavy ion collisions: 'Event by
  %event P and CP violation',''
  Nucl.\ Phys.\  A {\bf 803}, 227 (2008); K.~Fukushima, D.~E.~Kharzeev and H.~J.~Warringa,
  %``The Chiral Magnetic Effect,''
  Phys.\ Rev.\  D {\bf 78}, 074033 (2008).

\bibitem{STAR-CP}B.~I.~Abelev {\it et al.}  [STAR Collaboration],
  %``Azimuthal Charged-Particle Correlations and Possible Local Strong Parity
  %Violation,''
  Phys.\ Rev.\ Lett.\  {\bf 103}, 251601 (2009).

\bibitem{CME-alternate}J.~Liao, V.~Koch and A.~Bzdak,
  %``On the Charge Separation Effect in Relativistic Heavy Ion Collisions,''
  Phys.\ Rev.\  C {\bf 82}, 054902 (2010); S.~Schlichting and S.~Pratt,
  %``Charge conservation in RHIC and contributiuons to local parity violation
  %observables,''
  arXiv:1009.4283 [nucl-th]; F.~Wang,
  %``Effects of Cluster Particle Correlations on Local Parity Violation
  %Observables,''
  Phys.\ Rev.\  C {\bf 81}, 064902 (2010).

\bibitem{Fraga}E.~S.~Fraga, A.~J.~Mizher and M.~N.~Chernodub,
  %``Possible splitting of deconfinement and chiral transitions in strong
  %magnetic fields in QCD,''
  arXiv:1011.5626 [hep-ph].

\bibitem{Mrowczynski}S.~Mrowczynski and M.~H.~Thoma,
  %``What do electromagnetic plasmas tell us about quark-gluon plasma?,''
  Ann.\ Rev.\ Nucl.\ Part.\ Sci.\  {\bf 57}, 61 (2007).

\bibitem{cosmo-reviews}P.~B.~Greene, L.~Kofman, A.~D.~Linde and A.~A.~Starobinsky,
  %``Structure of resonance in preheating after inflation,''
  Phys.\ Rev.\  D {\bf 56}, 6175 (1997); R.~Micha and I.~I.~Tkachev,
  %``Turbulent thermalization,''
  Phys.\ Rev.\  D {\bf 70}, 043538 (2004).

\bibitem{Dusling-EGV}K.~Dusling, T.~Epelbaum, F.~Gelis and R.~Venugopalan,
  %``Role of quantum fluctuations in a system with strong fields: Onset of
  %hydrodynamical flow,''
  arXiv:1009.4363 [hep-ph].

\bibitem{Son-etal}P.~Kovtun, D.~T.~Son and A.~O.~Starinets,
  %``Viscosity in strongly interacting quantum field theories from black hole
  %physics,''
  Phys.\ Rev.\ Lett.\  {\bf 94}, 111601 (2005); G.~Policastro, D.~T.~Son and A.~O.~Starinets,
  %``The shear viscosity of strongly coupled N = 4 supersymmetric Yang-Mills
  %plasma,''
  Phys.\ Rev.\ Lett.\  {\bf 87}, 081601 (2001).

\bibitem{Myers}A.~Buchel, R.~C.~Myers and A.~Sinha,
  %``Beyond eta/s = 1/4pi,''
  JHEP {\bf 0903}, 084 (2009).

\bibitem{Maldacena}J.~M.~Maldacena,
  %``The large N limit of superconformal field theories and supergravity,''
  Adv.\ Theor.\ Math.\ Phys.\  {\bf 2}, 231 (1998).

\bibitem{visc-hydro-reviews}H.~Song and U.~W.~Heinz,
  %``Multiplicity scaling in ideal and viscous hydrodynamics,''
  Phys.\ Rev.\  C {\bf 78}, 024902 (2008); K.~Dusling and D.~Teaney,
  %``Simulating elliptic flow with viscous hydrodynamics,''
  Phys.\ Rev.\  C {\bf 77}, 034905 (2008); P.~Romatschke and U.~Romatschke,
  %``Viscosity Information from Relativistic Nuclear Collisions: How Perfect is
  %the Fluid Observed at RHIC?,''
  Phys.\ Rev.\ Lett.\  {\bf 99}, 172301 (2007); M.~Luzum and P.~Romatschke,
  %``Conformal Relativistic Viscous Hydrodynamics: Applications to RHIC results
  %at sqrt(s_NN) = 200 GeV,''
  Phys.\ Rev.\  C {\bf 78}, 034915 (2008)
  [Erratum-ibid.\  C {\bf 79}, 039903 (2009)]; B.~Schenke, S.~Jeon and C.~Gale,
  %``Elliptic and triangular flow in event-by-event (3+1)D viscous
  %hydrodynamics,''
  arXiv:1009.3244 [hep-ph].

\bibitem{Schafer-Teaney}T.~Schafer and D.~Teaney,
  %``Nearly Perfect Fluidity: From Cold Atomic Gases to Hot Quark Gluon
  %Plasmas,''
  Rept.\ Prog.\ Phys.\  {\bf 72}, 126001 (2009).

\bibitem{Moore}S.~C.~Huot, S.~Jeon and G.~D.~Moore,
  %``Shear viscosity in weakly coupled ${\cal N} {=} 4$ Super Yang-Mills theory
  %compared to QCD,''
  Phys.\ Rev.\ Lett.\  {\bf 98}, 172303 (2007).

\bibitem{Lacey}R.~A.~Lacey {\it et al.},
  %``Has the QCD critical point been signaled by observations at RHIC?,''
  Phys.\ Rev.\ Lett.\  {\bf 98}, 092301 (2007).

\bibitem{Alice-v2talk}Talk by J. Schukraft on behalf of the ALICE collaboration, CERN LPCC seminar, December 2, (2010).

\bibitem{Asakawa-BM}M.~Asakawa, S.~A.~Bass and B.~Muller,
  %``Anomalous viscosity of an expanding quark-gluon plasma,''
  Phys.\ Rev.\ Lett.\  {\bf 96}, 252301 (2006).

\bibitem{ALICE-v2}K.~Aamodt {\it et al.}  [The ALICE Collaboration],
  %``Elliptic flow of charged particles in Pb-Pb collisions at 2.76 TeV,''
  arXiv:1011.3914 [nucl-ex].

\bibitem{Luzum}M.~Luzum,
  %``Elliptic flow at LHC: Comparing heavy ion data to viscous hydrodynamic
  %prediction,''
  arXiv:1011.5173 [nucl-th]; G.~Kestin and U.~W.~Heinz,
  %``Hydrodynamic radial and elliptic flow in heavy-ion collisions from AGS to
  %LHC energies,''
  Eur.\ Phys.\ J.\  C {\bf 61}, 545 (2009).

\bibitem{Dumitru-MN}A.~Dumitru, E.~Molnar and Y.~Nara,
  %``Entropy production in high-energy heavy-ion collisions and the
  %correlation of shear viscosity and thermalization time,''
  Phys.\ Rev.\  C {\bf 76}, 024910 (2007)

\bibitem{Sorensen}P.~Sorensen,
  %``Implications of space-momentum correlations and geometric fluctuations in
  %heavy-ion collisions,''
  J.\ Phys.\ G {\bf 37}, 094011 (2010); R.~A.~Lacey, R.~Wei, N.~N.~Ajitanand and A.~Taranenko,
  %``Initial eccentricity fluctuations and their relation to higher-order flow
  %harmonics,''
  arXiv:1009.5230 [nucl-ex].

\bibitem{Ollitrault-talk}J.~Y.~Ollitrault,
  %``Phenomenology of the little bang,''
  arXiv:1008.3323 [nucl-th].

\bibitem{jetquenching}K.~Adcox {\it et al.}  [PHENIX Collaboration],
  %``Suppression of hadrons with large transverse momentum in central Au+Au
  %collisions at $\sqrt{s_{NN}}$ = 130-GeV,''
  Phys.\ Rev.\ Lett.\  {\bf 88}, 022301 (2002); C.~Adler {\it et al.}  [STAR Collaboration],
  %``Disappearance of back-to-back high $p_{T}$ hadron correlations in central
  %Au+Au collisions at $\sqrt{s_{NN}}$ = 200-GeV,''
  Phys.\ Rev.\ Lett.\  {\bf 90}, 082302 (2003).

\bibitem{dA}S.~S.~Adler {\it et al.}  [PHENIX Collaboration],
  %``Absence of suppression in particle production at large transverse  momentum
  %in s(NN)**(1/2) = 200-GeV d + Au collisions,''
  Phys.\ Rev.\ Lett.\  {\bf 91}, 072303 (2003); .~Adams {\it et al.}  [STAR Collaboration],
  %``Evidence from d + Au measurements for final-state suppression of high  p(T)
  %hadrons in Au + Au collisions at RHIC,''
  Phys.\ Rev.\ Lett.\  {\bf 91}, 072304 (2003).

\bibitem{Bj-Appel}J.~D.~Bjorken,
  %``Energy Loss Of Energetic Partons In Quark - Gluon Plasma: Possible
  %Extinction Of High P(T) Jets In Hadron - Hadron Collisions,''
  FERMILAB-PUB-82-059-THY; D.~A.~Appel,
  %``Jets As A Probe Of Quark - Gluon Plasmas,''
  Phys.\ Rev.\  D {\bf 33}, 717 (1986).

\bibitem{Gyulassy-Wang}M.~Gyulassy and X.~n.~Wang,
  %``Multiple collisions and induced gluon Bremsstrahlung in QCD,''
  Nucl.\ Phys.\  B {\bf 420}, 583 (1994)

\bibitem{BDMPS}R.~Baier, Y.~L.~Dokshitzer, A.~H.~Mueller, S.~Peigne and D.~Schiff,
  %``Radiative energy loss and p(T)-broadening of high energy partons in
  %nuclei,''
  Nucl.\ Phys.\  B {\bf 484}, 265 (1997).

\bibitem{quenching-reviews}U.~A.~Wiedemann,
  %``Jet Quenching in Heavy Ion Collisions,''
  arXiv:0908.2306 [hep-ph]; A.~Majumder and M.~Van Leeuwen,
  %``The theory and phenomenology of perturbative QCD based jet quenching,''
  arXiv:1002.2206 [hep-ph]; D.~d'Enterria,
  %``Jet quenching,''
  arXiv:0902.2011 [nucl-ex].

\bibitem{Gale-Huot}S.~Caron-Huot and C.~Gale,
  %``Finite-size effects on the radiative energy loss of a fast parton in hot
  %and dense strongly interacting matter,''
  arXiv:1006.2379 [hep-ph].

\bibitem{ALICE-jetquenching}K.~Aamodt {\it et al.}  [ALICE Collaboration],
  %``Suppression of Charged Particle Production at Large Transverse Momentum in
  %Central Pb--Pb Collisions at $\sqrt{s_{_{NN}}} = 2.76$ TeV,''
  arXiv:1012.1004 [nucl-ex].

\bibitem{Paukkunen}H.Paukkunen, these proceedings; K.~J.~Eskola, H.~Paukkunen and C.~A.~Salgado,
  %``EPS09 - a New Generation of NLO and LO Nuclear Parton Distribution
  %Functions,''
  JHEP {\bf 0904}, 065 (2009).

\bibitem{Iancu-review}E.~Iancu,
  %``Partons and jets in a strongly-coupled plasma from AdS/CFT,''
  Acta Phys.\ Polon.\  B {\bf 39}, 3213 (2008).

\bibitem{Liu-etal}H.~Liu, K.~Rajagopal and U.~A.~Wiedemann,
  %``Calculating the jet quenching parameter from AdS/CFT,''
  Phys.\ Rev.\ Lett.\  {\bf 97}, 182301 (2006).

\bibitem{Yaffe-Gubser-Teaney}J.~Casalderrey-Solana and D.~Teaney,
  %``Heavy quark diffusion in strongly coupled {\cal N} = 4 Yang-Mills,''
  Phys.\ Rev.\  D {\bf 74}, 085012 (2006); S.~S.~Gubser,
  %``Momentum fluctuations of heavy quarks in the gauge-string duality,''
  Nucl.\ Phys.\  B {\bf 790}, 175 (2008); P.~M.~Chesler and L.~G.~Yaffe,
  %``The stress-energy tensor of a quark moving through a strongly-coupled N=4
  %supersymmetric Yang-Mills plasma: comparing hydrodynamics and AdS/CFT,''
  Phys.\ Rev.\  D {\bf 78}, 045013 (2008).

\bibitem{Pisarski}
  R.~D.~Pisarski,
  %``Quark-gluon plasma as a condensate of SU(3) Wilson lines,''
  Phys.\ Rev.\  D {\bf 62}, 111501 (2000).

\bibitem{Caron-huot}S.~Caron-Huot,
  %``O(g) plasma effects in jet quenching,''
  Phys.\ Rev.\  D {\bf 79}, 065039 (2009); P.~B.~Arnold and W.~Xiao,
  %``High-energy jet quenching in weakly-coupled quark-gluon plasmas,''
  Phys.\ Rev.\  D {\bf 78}, 125008 (2008).

\bibitem{ATLAS-jet}ATLAS Collaboration,
  %``Observation of a Centrality-Dependent Dijet Asymmetry in Lead-Lead
  %Collisions at sqrt(S(NN))= 2.76 TeV with the ATLAS Detector at the LHC,''
  arXiv:1011.6182 [hep-ex].

\bibitem{CMS-jet}Talk by B. Wyslouch on behalf of the CMS collaboration, CERN LPCC seminar, December 2, (2010).

\bibitem{Solana-MW}J.~Casalderrey-Solana, J.~G.~Milhano and U.~A.~Wiedemann,
  %``Jet Quenching via Jet Collimation,''
  arXiv:1012.0745 [hep-ph].


\end{thebibliography}
\end{document}